\documentclass[a4paper]{article}
\usepackage{amsmath,amssymb,amsthm,fullpage}
\usepackage[shortcuts]{extdash}
\usepackage{hyperref}

\usepackage{csquotes}
\usepackage[backend=biber,maxbibnames=99,sorting=none]{biblatex}
\bibliography{cymoduli.bib}

\usepackage{color}
\usepackage{lipsum}
\usepackage{amssymb,amsmath,amsfonts,epic,eepic,float}

\usepackage[mathscr]{eucal}

\usepackage{rotating,epsfig,indentfirst,array,varioref}
\usepackage{appendix,marginnote,tikz,pgf,mathtools}
\usepackage{setspace}

\usepackage{authblk}

\makeatletter
\def\@@bfil{\leaders \vrule \@height \ht\z@ \@depth \z@ \hfill}% default brace filler
\def\@bLfil{\@@bfil}% left leader filler
\def\@bRfil{\@@bfil}% right leader filler
\def\resetbraceratio{\gdef\@bLfil{\@@bfil}\gdef\@bRfil{\@@bfil}}% reset to default braces
\def\setbraceratio#1#2{% \setbraceratio{<left>}{<right>}
  \let\@bLfil\relax% clear left filler
  \multido{\iA=1+1}{#1}{\gappto\@bLfil{\@@bfil}}% increase left ratio
  \let\@bRfil\relax% clear right filler
  \multido{\iA=1+1}{#2}{\gappto\@bRfil{\@@bfil}}% increase right ratio
}
\def\upbracefill{$\m@th\setbox\z@\hbox{$\braceld$}\bracelu\@bLfil\bracerd\braceld\@bRfil\braceru$}
\def\downbracefill{$\m@th\setbox\z@\hbox{$\braceld$}\braceld\@bLfil\braceru\bracelu\@bRfil\bracerd$}
\makeatother

\usepackage{tikz,pgf}
\usetikzlibrary{shapes}
\usetikzlibrary{calc}
\usetikzlibrary{decorations.pathmorphing}
\usetikzlibrary{decorations.pathreplacing,shapes.misc}
\usetikzlibrary{positioning}
\usetikzlibrary{arrows}
\usetikzlibrary{decorations.markings}

\def\be{\begin{equation}}
\def\ee{\end{equation}}
\def\barr{\begin{array}}
\def\earr{\end{array}}

\def\1{\tilde{1}}
\def\2{\tilde{2}}
\def\3{\tilde{3}}

\newcommand{\ba}{\begin{equation}\begin{aligned}}
\newcommand{\ea}{\end{aligned}\end{equation}}

\newcommand{\bml}{\begin{multline}}
\newcommand{\eml}{\end{multline}}

\newcommand{\CC}{\mathbb{C}}
\newcommand{\RR}{\mathbb{R}}

\newcommand{\ZZ}{\mathbb{Z}}
\newcommand{\PP}{\mathbb{P}}

\newcommand{\dd}{\mathrm{d}}
\newcommand{\pd}{\partial}

\newcommand{\Res}{\mathrm{Res}}

\begin{document}

\title {Exact computation of the Special geometry for Calabi--Yau hypersurfaces of Fermat type.}

\author{Konstantin Aleshkin $^{1,2}$\thanks{kaleshkin@sissa.it},}
\author{Alexander Belavin $^{1,3,4}$\thanks{belavin@itp.ac.ru} }

\affil{$^1$ L.D. Landau Institute for Theoretical Physics\\
 Akademika Semenova av. 1-A\\ Chernogolovka, 142432  Moscow region, Russia}

\affil{$^2$ International School of Advanced Studies (SISSA),
 via Bonomea 265, 34136 Trieste, Italy}
\affil{$^3$ Moscow Institute of Physics and Technology\\
Dolgoprudnyi, 141700 Moscow region, Russia}
\affil{$^4$ Institute for Information Transmission Problems\\
 Bolshoy Karetny per. 19, build.1, Moscow 127051 Russia}

\maketitle
\abstract{We continue to develop our method for effectively computating the special K\"ahler geometry on the moduli space of Calabi--Yau manifolds. 
We generalize it to all polynomial deformations of Fermat hypersurfaces.}

\flushbottom

%%%%%%%%%%%%%%%%%%%%%%%%%%%%%%%%%%%%%%%%%%%%%%%%%%%%%%%%%%%%%%%%%%%%%%%%%%%%%%%%%%%%%%%%%%%%%%%%%%%%%%%%%%%%
 \section{Introduction.}
 When the Superstring theory is  compactified on a Calabi--Yau (CY) threefold $X,$ 
the low-energy effective theory is defined  in terms of the  Special K\"ahler  geometry of the CY moduli space.\\
\indent       It is known that  the K\"ahler potential is given by the logarithm of the holomorphic
 volume of CY manifold~\cite{Distances,Rolling}  $X_{\phi}$:
\ba
G(\phi)_{a\bar{b}} &= \pd_a \overline{\pd_b} \, K(\phi, \bar{\phi}), \\
e^{-K(\phi)} &= \int_{X_{\phi}} \Omega\wedge\overline{\Omega},
\ea
This can be rewritten  in terms of periods of  $\Omega$ as
\ba
\omega_{\mu}(\phi) &:= \int_{q_{\mu}} \Omega, \;\;\; q_{\mu} \in H_3(X, \RR), \\
e^{-K} &= \omega_{\mu}(\phi) C_{\mu\nu}\; \overline{\omega_{\nu}(\phi)},
\ea
where  $C_{\mu\nu} = [q_{\mu}] \cap [q_{\nu}]$ is an intersection matrix of 
3-cycles. The K\"ahler metric on the moduli space is also called as the Weil--Petersson
metric or $tt^*$ metric~\cite{CV} and is closely related to the  Zamolodchikov metric.
Apart from its own interest, the knowledge of its explicit form is useful in various
 contexts. In particular it enters the vacua equation in the moduli stabilization
problem~\cite{QuevedoEtAl} and the  holomorphic anomaly equation~\cite{BCOV} for the higher
genus B-model and allows  computing distances in the moduli space. For instance, 
 it was recently used to check the Refined Swampland Distance Conjecture 
in~\cite{BlumenhagenEtAl, Blumenhagen}.
This metric  was first found for the Quintic threefold in~\cite{COGP}.
There were then results in the one-modulus cases in~\cite{Klemm} and in some
2-moduli cases~\cite{CFKM, COFKM}. As far as we know, there have not been many
 advances apart from these until recently. Another interesting
 approach to computing the Special
geometry was proposed in~\cite{JockersMorrison} based on mirror symmetry and
 localization computations
in 2-dimensional models~\cite{BeniniCremonesi, DoroudGomis}.

We lately suggested a new much simpler  method for computing the K\"ahler metric 
for a large class of CY  defined as hypersurfaces in weighted projective spaces \cite {AKBA,AKBA2,AKBA3}. 
This method uses the correspondence between the middle cohomology of CY manifolds  and
 the invariant Frobenius algebra associated  with the potential $W$  defining the given CY manifold.
This correspondence  is realized by  the oscillatory integral representation  for the periods of the holomorphic CY 3-form.
Having established this correspondence we obtain an efficient method 
for computing the special geometry on the moduli space. The important feature of 
this method is that it allows to compute the K\"ahler potential for all polynomial
deformations of the CY manifold as opposed to a few moduli cases before.

If a CY manifold $X$ is realized by a  quasi-homogeneous polynomial  $W(x)$ in a weighted projective space $\PP_{k_1,...,k_5}^4$, then a subgroup of
the cohomology group  $H^3(X)$  with its natural Hodge decomposition 
$H^3(X) = H^{3,0}(X)\oplus H^{2,1}(X)\oplus H^{1,2}(X)\oplus H^{0,3}(X)$,
 the complex conjugation $*$  of differential forms and Poincare pairing $\langle \cdot
, \cdot\rangle$ is isomorphic
to  the invariant Milnor ring $R^Q$ defined by  $W(x)$  with a Hodge decomposition
 given by the monomial weight grading, antiholomorphic involution $M$ and the
 residue pairing  $ \eta$. 
From this fact, we obtain the formula for the K\"ahler potential $K(\phi)$
\be \label{form}
e^{-K(\phi)} = \sigma^+_{\mu}(\phi)\; \eta_{\mu\lambda}\; M_{\lambda \nu}\;
 \overline{\sigma^-_{\nu}(\phi)},
\ee
where $\sigma^{\pm}_{\mu}(\phi)$ are periods  computed as oscillatory  integrals,
$\eta_{\mu\nu}$ is a residue pairing in the Milnor ring and 
 $M_{\mu \nu}$ is  the   antiholomorphic involution of the  ring $R^Q$. 
The three ingredients $\sigma_{\mu}(\phi),\;\eta_{\mu\nu}$ and $\;M_{\mu \nu}$ can all be efficiently computed.

For CY manifolds defined as a Fermat hypersurface we explicitly compute the K\"ahler potential as a function of \textit{all polynomial deformation parameters}. The main
result of this paper is presented in section~\ref{sec:res}.
The computation is not more difficult then
in the Quintic threefold case. Here we compute this case and find
the real structure (see below) more transparently and simply.
For Fermat polynomials, we can compute the real structure matrix $M$ using decomposition
of the oscillatory integrals into the product of one-dimensional ones.

Section~\ref{reminder} of the paper is a short exposition of the methods developed
in~\cite{AKBA, AKBA2, AKBA3} and section~\ref{computation} is a specification and
computation for the Fermat case.

%%%%%%%%%%%%%%%%%%%%%%%%%%%%%%%%%%%%%%%%%%%%%%%
%%%%%%%%%%%%%%%%%%%%%%%%%%%%%%%%%%%%%%%%%%%%%%
%%%%%%%%%%%%%%%%%%%%%%%%%%%%%%%%%%%%%%%%%%%%%%%%%%%%%%%%%%%%%%%%%%%%%%%%%%%%%%%%%%%%%%%
%%%%%%%%%%%%%%%%%%%%%%%%%%%%%%%%%%%%%%%%%%%%%%%%%%%%%%%%%%%%%%%%%%%%%%%%%%%%%%%%%%%%%%%
 \section{CY as the hypersurface in the weighted projective space} \label{reminder}
  Let 
\be
\PP^{4}_{(k_1, \ldots, k_5)} = \{(x_1 : \cdots, x_5) \; | \; (x_1 : \cdots, x_5) \simeq
 ( \lambda^{k_1} x_1 : \cdots, \lambda^{k_5} x_5)\}
\ee
 denote the weighted projective space\footnote{Although  these spaces in general are orbifolds, we work with them as if they were smooth.  For our purposes  this does not lead to any problems in 
computing  of the integrals over 3-cycles.} and
\be
X = \{ x_1, \ldots, x_5 \; \in \PP^{4}_{(k_1, \ldots, k_5)} | \; W_0(x) = 0 \}
\ee
be a transverse hypersurface for some quasi-homogeneous polynomial  $W_0(x)$, 
\be
W_0(\lambda^{k_i} x_i) = \lambda^d W_0(x_i)
\ee
  and 
\be
\operatorname{deg} W_0(x) = d = \sum_{i=1}^5 k_i.
\ee
The last relation ensures that $X$ is a CY manifold.
The (\textit{polynomial} part of the) moduli space of complex structures of $X$ is then  given by  homogeneous polynomial
 deformations of this singularity up to coordinate transformations:
\be
W(x, \phi) = W_0(x) + \sum_{s}^{h^{\mathrm{poly}}_{21}} \phi_{{\bf{s}}}  e_s(x),
\ee
where $e_s(x)$ are monomials in $x$ of the same degree  $d$.  We use the index $s$ for monomials and periods in this range.
The holomorphic 3-form $\Omega$ is given as a residue of a 5-form in the underlying
 affine space $\CC^5$:
$$
\Omega = \frac{x_5 \dd x_1\wedge \dd x_2\wedge \dd x_{3}}{\pd W(x)/\pd x_{4}}.
$$
We define period integrals or periods of $\Omega$ needed for our goal:
$$
\omega_{\mu}(\phi) := \int_{q_{\mu}} \Omega, \;\;\; q_{\mu} \in H_3(X, \RR),
$$
where $H^3(X)$ has a natural Hodge structure $H^3(X) = \oplus_{k=0}^{3} H^{3-k, k}(X)$,
$$\dim H^{3, 0}(X)=\dim H^{0,3}(X)=1,\;\dim H^{2, 1}(X)=\dim H^{1,2}(X)=h^{2,1}.$$ \\
The Poincar\'e pairing can be written in terms of  integrals over some
 cycles $q_{\mu}$ as
$$
%\be \label{poinc}
\eta(\chi_a, \chi_b) = \int_X \chi_a \wedge \chi_b = \int_{q_{\mu}} \chi_a \; C_{\mu\nu} \int_{q_{\nu}} \chi_b
$$
%\ee
and is invariant under complex conjugation $(p,q)$-forms. Here, 
$C_{\mu\nu} = [q_{\mu}] \cap [q_{\nu}]$ is the intersection matrix of  3-cycles.
%%%%%%%%%%%%%%%%%%%%%%%%%%%%%%%%%%%%%%%%%%%%%
%%%%%%%%%%%%%%%%%%%%%%%%%%%%%%%%%%%%%%%%%%%%
\subsection{Q-invariant Milnor ring.}
On the other hand, the polynomial $W_0(x)$, considered as a singularity,
 defines a Milnor ring~\cite{AVG} $R_0$.
Let $Q= \ZZ_d$ denote a symmetry group $Q$
 acting on $\CC^5$ diagonally as $\lambda \cdot (x_1, \cdots , x_5) = (\lambda^{k_1} 
x_1, \cdots ,\lambda^{k_5} x_5)$ for $\lambda^d = 1$. This action is trivial on
 $\PP^4_{(k_1, \cdots, k_5)}$ and, moreover,  preserves $W(x, \phi)$.
We consider the $Q$-invariant part of the Milnor ring
\ba
R^Q = \left(\frac{\CC[x_1, \ldots, x_5]}{\mathrm{Jac}(W_0)}\right)^Q,\quad
\mathrm{Jac}(W_0) = \langle \pd_i W_0 \rangle_{i=1}^5.
\ea
The subring $R^Q$ becomes  a Frobenius ring if it is endowed     with  the  pairing 
$$\eta(e_{\alpha}, e_{\beta}) =\Res \frac{e_{\alpha}(x) e_{\beta}(x) \, \dd^5 x}
{\prod_{i=1}^N \pd_i W_0(x)}.$$
 The Hodge decomposition of $R^Q $   
  corresponds  with  to the quasi-homogeneity degrees $0,d,2d,3d$ of its components 
 $$R^Q = (R^Q)^0 \oplus (R^Q)^1 \oplus (R^Q)^2 \oplus (R^Q)^3,$$
  where 
$\dim (R^Q)^0 =\dim (R^Q)^3=1,\; \; $ $\dim (R^Q)^1 =\dim  
(R^Q)^2=h_{2,1}$,
and $\dim\; R^Q =\dim H^3(X)$. 
Here $h_{2,1}$ denotes the
 number of deformations of complex struture which can be represented as
 polynomial deformations in the ambient projective space.
In particular, $(R^Q)^3 = \langle e_{\rho}(x) \rangle$, where we introduced  the notation
\be
e_{\rho}(x) = \det \pd_i \pd_j W_0(x).
\ee
%%%%%%%%%%%%%%%%%%%%%%%%%%%%%%%%%%%%%%%%%%%%%
%%%%%%%%%%%%%%%%%%%%%%%%%%%%%%%%%%%%%%%%%%%%
\subsection{$\bf{Q}$-invariant cohomology $\bf{H^5_{D_{\pm}}(\CC^5)_{inv}}$ and the  oscillatory integrals.}
In the next step, we define two  differentials $D_{\pm}$,
 \be
D_{\pm} = e^{\mp W_0} \dd \, e^{\pm W_0} = \dd \pm \dd W_0\wedge,  \qquad ( D_{\pm})^2 =0,
\ee
and  two groups of $Q$-invariant cohomology $H^5_{D_{\pm}}(\CC^5)_{Q}$. As vector spaces,
they are isomorphic to $R^Q$.
Choosing  some basis   $\{e_{\mu}(x) \}$  in the ring  $R^Q$ we can write a basis of
 $H^5_{D_{\pm}}(\CC^5)_{Q}$ as $\{e_{\mu}(x) \, \dd^5 x\}$.
These groups inherit the grading degree and Hodge structure from $R^Q$.
These cohomology groups are naturally subgroups of the middle cohomology group
 $\in H^3(X)$ (\cite{Candelas}).
 This isomorphism, defined below, maps the Hodge decomposition components
 of $H^5_{\pm}(\CC^5)_{Q}$
 spanned by $e_{\mu}(x) \, \dd^5 x$ with $e_{\mu}(x) \in (R^Q)^{q}$ to
the corresponding components $H^{3-q,q}(X)$. 
 It also sends the Poincare pairing on the differential forms on $X$
to the invariant  ring $R^Q$  pairing $\eta$. 
%%%%%%%%%%%%%%%%%%%%%%%%%%%%%%%%%%%%%%%%%%%%%%%%%%%%%%%%%%%%%%%%%%%%%%%%%%%%%%%%%%
Having $H^5_{D_{\pm}}(\CC^5)_{Q}$, we define their dual homology group, i. e.,
the  Q-invariant relative homology groups
$\mathscr{H}_{5}^{\pm,Q} := H_5(\CC^5, \;\mathrm{Re}W_0=L \to \pm \infty)_{Q}$ 
as a quotient of the relative homology group
$H_5(\CC^5,\;\mathrm{Re}W_0=L \to \pm \infty)$.
 For this,  we define the pairing via oscillatory integrals
\be
\langle e_{\mu}(x)\, \dd^5 x, Q^{\pm}_{\nu} \rangle: = \int_{Q^{\pm}_{\nu}} 
e_{\mu}(x) \, e^{\mp W(x)} \dd^5 x. 
\ee
Using  this pairing, we define  the relative invariant homology  groups $\mathscr{H}_{5}^{\pm,Q}$  as the quotient of  $ H_5(\CC^5, W_0 = L,\;\mathrm{Re}L \to \pm \infty)$  by its subspace whose elements are orthogonal to  all elements of    $H^5_{D_{\pm}}(\CC^5)_{Q}$.
 The crucial fact in what follows  is   that  $R^Q$ and $H^3(X)$ 
and all  their additional structures  are isomorphic to each other.
First, there exists   an isomorphism $S$ of cycles  for each $\phi$.  This gives
\be
S(Q^+_{\mu}) = q_{\mu},\; \; \; Q^+_{\mu}\in \mathscr{H}_{5}^{\pm,Q}, \;\; q_{\mu}\in  H_3(X, \ZZ).
\ee
The isomorphism is defined  by the oscilatory integrals as follows.
Let $\{q_{\mu}\}$ is a basis  of $H_{3}(X, \ZZ),$ then  the  basis
$Q^{\pm}_{\mu}$ of $\mathscr{H}_{5}^{\pm,Q}$ can be choosen
in such a way that the integrals over the corresponding cycles of these bases are equal
\be \label{iso1}
\int_{q_{\mu}} \Omega_{\phi} = \int_{Q_{\mu}^{\pm}} e^{\mp W(x, \phi)} \, \dd^5 x.
\ee
%%%%%%%%%%%%%%%%%%%%%%%%%%%%%%%%%%%%%%%%%%%%%%%%%
%%%%%%%%%%%%%%%%%%%%%%%%%%%%%%%%%%%%%%%%%%%%%%%%%%%%%%%%%%%%%%%%%%%%%%%%%%%%%%%%%%%%%%%
%%%%%%%%%%%%%%%%%%%%%%%%%%%%%%%%%%%%%%%%%%%%%%%%
\subsection{ $\bf{ H^3(X)}$  versus    $\bf{H^5_{D_{\pm}}(\CC^5)_{\mathrm{inv}}}$ correspondence.}

Having an  isomorphism between $H_3(X)$ and  $\mathscr{H}_{5}^{\pm,Q}$,  we   define
the isomorphism between  the two  cohomology groups
 $H^3(X)$    and  $H^5_{D_{\pm}}(\CC^5)_{Q}$ also using oscillatory integrals.
We take the basis of  cycles $q_{\mu}\in H_3(X)$ and   the corresponding   basis of   cycles  
$Q^{\pm}_{\mu} \in \mathscr{H}_{5}^{\pm,Q}$  at $\phi=0$.
The  form  $\chi_{\alpha}\in H^3(X)$ then    corresponds to 
the   form $e_{\alpha}(x) \, \dd^5 x \in H^5_{D_{\pm}}(\CC^5)_{Q}$  iff
\ba \label{inteq}
\int_{q_{\mu}} \chi_{\alpha} = \int_{Q_{\mu}^{\pm}} e_{\alpha}(x) \, 
 e^{\mp W(x, \phi)} \, \dd^5 x \;\;\; 
\ea
for all pairs $\{q_{\mu},Q_{\mu}\}$.
Thus these  two forms are isomorphic if they have equal coordinates  (i. e., periods) 
in  some isomorphic bases.\\
This isomorphism preserves  the Hodge \textit{filtration} i. e.,
the elements $(R_0^Q)^{\le k d}$ are mapped to $F^kH^3(X) := \oplus_{i \le k} H^{3-i,i}(X)$.
 This can been  seen by 
 differentiating~\eqref{iso1} with respect to the deformation parameters $\phi$. The kth derivative of the
RHS belongs to $\oplus_{i \le k} H^{3-i,i}(X)$ by Kodaira's lemma or Griffiths transversality
while the kth derivative of the LHS belongs to $(R_0^Q)^{\le k d}$. As is seen below,
for Fermat hypersurfaces, the isomorphism also  preserves  the decomposition (therefore
kth derivative of the RHS belongs to $H^{3-k,k}(X)$.
The intersection matrices of the  cycles $q_{\mu} \cap q_{\nu}$ and $Q^+_{\mu} \cap
Q^-_{\nu}$  coincide, as we now show.
It follows from  the coincidence of   the pairings of the differential  forms $\in H^3(X)$ and of  the corresponding  elements $\in R^Q$. 

\indent We rewrite  the Poincar\'e pairing  of  $\chi_a, \chi_b$ in   $H^3(X),$  
\be
\langle\chi_a, \chi_b\rangle: = \int_X \chi_a \wedge \chi_b
\ee
 as the bilinear expression  of periods,
\be
\langle\chi_a, \chi_b\rangle = \int_{q_{\mu}} \chi_a \; C_{\mu\nu} \int_{q_{\nu}} \chi_b,
\ee
where $C_{\mu\nu} = q_{\mu} \cap q_{\nu}$ is the intersection matrix of the cycles.
On the other hand, the residue pairing $\eta(e_{a}, e_{b})$ in the ring $R^Q$  can be written
in terms of  the periods as explained in~\cite{Cecotti,CV}, also see~\cite{Chiodo}
  as
\be
\eta(e_{a}, e_{b}) = \int_{Q^+_{\mu}} e_a e^{-W_0(x)} \dd^5 x \;
\hat {C}_{\mu\nu}  \int_{Q^-_{\mu}} e_b e^{W_0(x)} \dd^5 x,
\ee
where $\hat{C}_{\mu\nu} = Q^+_{\mu} \cap Q^-_{\nu}$.
Taking   into account  the  eguality~\eqref{inteq} and the equality of the pairings 
\be
 \langle\chi_a, \chi_b\rangle=\eta(e_{a}, e_{b}) 
\ee
we obtain  the relation  $\hat{C}_{\mu\nu}={C}_{\mu\nu}$.
The  relation (18) will be   used below  for  expressing the intersection matrix $ C_{\mu\nu}$ 
 in terms of  the $ R^Q$ pairing.
%%%%%%%%%%%%%%%%%%%%%%%%%%%%%%%%%%%%%%%%%%%%%%%%%%%%%%%%%%%%%%%%%%%%%%%%%%%%%%%%%%%%%%%
%%%%%%%%%%%%%%%%%%%%%%%%%%%%%%%%%%%%%%%%%%%%%%%%%%%%%%%%%%%%%%%%%%%%%%%%%%%%%%%%%%%%%%%
\subsection{Anti-involution $M$ on $R^Q$}

The same isomorphism allows defining an anti-Involution $M$ 
on $R^Q$ and on the Q-invariant cohomology $H^5_{D_{\pm}}(\CC^5)_{inv}$
that corresponds to a complex conjugation $*$ on the differential forms in $H^3(X)$.
Let the form $\phi_{\mu} \in H^3(X)$  correspond to $\{e_{\mu}(x)\} \in R^Q$ under the
isomorphism $S$, and let
$$ *\phi_{\mu}= M_{\nu\mu} \phi_{\nu}. $$
Then   $R^Q$  inherits this involution. 
For the basis $\{e_{\mu}(x)\}$,  the antiholomorphic operation $*$  is
\be \label{conj}
*e_{\mu}(x) =  M_{\nu\mu} e_{\nu} (x).
\ee 
Because $(*)^2=I $, it follows from this definition  that $\bar M M=I $.\\
\indent We introduce  the convenient basis $\Gamma^{\pm}_{\mu}\in H^5_{D_{\pm}}(\CC^5)_{inv}$
dual to the basis $\{e_{\mu}(x)\}$  such that
\be
\langle  \Gamma^{\pm}_{\mu},e_{\nu}(x)\, \dd^5 x \rangle =
\int_{\Gamma^{\pm}_{\mu}}
e_{\nu}(x) \, e^{\mp W_0(x)} \dd^5 x   =\delta_{\mu\nu}. 
\ee
%%%%%%%%%%%%%%%%%%%%%%%%%%%%%%%%%%%%%%%%%%%%%%%%%
%%%%%%%%%%%%%%%%%%%%%%%%%%%%%%%%%%%%%%%%%%%%%%%%
This definition induces the antiholomorphic operation $*$  on  $\Gamma^{\pm}_{\mu}$
$$ *\Gamma^{\pm}_{\mu}= \bar{M}_{\mu\nu} \Gamma^{\pm}_{\nu},$$
and hence 
\be
\overline{\langle *\Gamma_{\mu}, e_{\nu}(x)\, \dd^5 x \rangle} = 
\langle \Gamma_{\mu}, *e_{\nu}(x)\, \dd^5 x \rangle.
\ee
The cycles $\Gamma^{\pm}_{\mu}$ belong to the homology group $\mathscr{H}_{5}^{\pm,Q}$, 
therefore  they are linear combinations of some geometric cycles with complex coefficients.
If we define $T$ as a transition  matrix from cycles  $\Gamma^{\pm}_{\mu}$ 
to an arbitrary real basis of cycles, for example,   Lefschetz thimbles $L^{\pm}_{\mu}=*L^{\pm}_{\mu}$, 
$$ L^{\pm}_{\mu}=T_{\mu \nu} \Gamma^{\pm}_{\nu},$$
then we have
$$ L^{\pm}_{\mu}=\overline{ T}_{\mu \nu}\; *\Gamma^{\pm}_{\nu}.$$
Comparing this relation with
$$ *\Gamma^{\pm}_{\mu}=\bar{M}_{\mu\nu} \Gamma^{\pm}_{\nu},$$
we obtain the  expression  for $M$ in terms of $T$,
$$M = T^{-1} \bar{T}.$$ 
Obviously  $M$ is  independent from the choice of real cycles.
From the definition  of the cycles $ \Gamma^{\pm}_{\mu}$, we obtain
   the useful relation   for computing $T_{\mu \nu}$ and $M_{\mu \nu}$  (as is seen below)
$$
T_{\mu \nu} =\int_{L^{\pm}_\mu} e_{\nu}(x) \, e^{\mp W_0(x)} \dd^5 x.
$$

%%%%%%%%%%%%%%%%%%%%%%%%%%%%%%%%%%%%%%%%%%%%%%%%%%%%%%%%%%%%%%%%%%%%%%%%%%%%%%%%%%%%%%%
%%%%%%%%%%%%%%%%%%%%%%%%%%%%%%%%%%%%%%%%%%%%%%%%%%%%%%%%%%%%%%%%%%%%%%%%%%%%%%%%%%%%%%%
\subsection{Deriving the main formula for K\"ahler potential }
The expression for the pairing on the ring $R^Q$  in terms of  periods  is
\be
\eta(e_{\mu}, e_{\nu}) = \int_{L^+_a} e_{\mu} e^{-W_0(x)} \dd^5 x \;
 C_{ab}  \int_{L^-_b} e_{\nu} e^{W_0(x)} \dd^5 x=T_{a \mu}C_{ab}T_{b \nu}.
\ee
We have also the formula for $K(\phi)$
$$
e^{-K} = {\omega^{+}_{b}(\phi)}\; C_{ab} \; \overline{\omega^{-}_{b}(\phi)}
$$
with
$$
\omega^{\pm}_{a}(\phi)=\int_{L^{\pm}_a} e^{\mp W(x,\phi)} \dd^5 x = T_{a\mu} \sigma^{\pm}_{\mu}(\phi),
$$
where  the periods $ \sigma^{\pm}_{\mu}(\phi)$ are  integrals  over cycles $\Gamma^{\pm}_{\mu}$
$$
 \sigma^{\pm}_{\mu}(\phi) =\int_{\Gamma^{\pm}_{\mu}} \, e^{\mp W(x,\phi)} \dd^5 x.
$$ 
Eliminating  the matrix $ C_{ab}$ from these relations   we obtain
\be 
e^{-K(\phi)} = \sum_{\mu,\nu, \lambda}  \sigma^+_{\mu}(\phi) \; \eta_{\mu\lambda} \;
 M_{\lambda \nu} \; \overline{\sigma^-_{\nu}(\phi)}.
\ee
%%%%%%%%%%%%%%%%%%%%%%%%%%%%%%%%%%%%%%%%%%%%%%%%%%%%%%%%%%%%%%%%%%%%%%%%%%%%%%%%%%%%%%%%%%%%%%%%%%%%%%%%%%%%
%%%%%%%%%%%%%%%%%%%%%%%%%%%%%%%%%%%%%%%%%%%%%%%%%%%%%%%%%%%%%%%%%%%%%%%%%%%%%%%%%%%%%%%
 \section{ Fermat threefolds} \label{computation}
 Let $X$ be a Fermat CY  $X = \{ x_1, \ldots, x_5 \; \in \PP^{4}_{(k_1, \ldots, k_5)} | \; W(x,\phi) = 0 \},$ 

$$W(x, \phi) = \sum_{i=1}^{5}x_i^{ \frac{d}{k_i}} + 
\sum_{\bf{s}=1}^{h^{\mathrm{poly}}_{21}} \phi_{{\bf{s}}}e_s(x), \quad  d = \sum_{i=1}^5 k_i,$$ 
where 
$\frac{d}{k_i}$ are positive  integers. 
The monomials  $e_s(x)=e_{(s_1,\cdots, s_5)}(x):=\prod_i x_i^{s_i}$ correspond to  deformations of the complex structure of $X$.
 Their weights are equal to $d$, 
$\sum_{i=1}^5 k_i s_i=d$, and 
each variable $x_i$ has a nonnegative integer power $s_i\le  \frac{d}{k_i}-2$.
The number  of such monomials is denoted $h^{\mathrm{poly}}_{21}$ and is less than
 or equal to the Hodge number 
$h_{21}$, and $\dim H_3(X) = 2 h_{21}+2$ (which can be verifed from the combinatorics
of the corresponding weighted projective space).
%We let $e_0(x) := e_{(1,1,1,1,1)}(x) = x_1x_2x_3x_4x_5$ to be the so-called
% fundamental monomial, which is somewhat distinguished in our picture. 
%%%%%%%%%%%%%%%%%%%%%%%%%%%%%%%%%%%%%%%%%%%%%%%%%%%%%%%%%%%%%%%%%%%%%%%%%%%%%%%%%%%%%%%
%%%%%%%%%%%%%%%%%%%%%%%%%%%%%%%%%%%%%%%%%%%%%%%%%%%%%%%%%%%%%%%%%%%%%%%%%%%%%%%%%%%%%%%
 \subsection{Q-invariant ring and phase symmetry.}

 The Milnor ring $R_0$ of the  Fermat polynomial $W_0(x)$ is generated as a vector space
 by monomials   $e_{\mu}(x)=\prod_i x_i^{\mu_i}$, where each nonnegative variable
 $\mu_i$ is less than $\frac{d}{k_i}-1$, and  $\dim R_0 =\prod_i (\frac{d}{k_i}-1) $.\\
\indent $R^Q$  is multiplicatively generated by the monomials $e_s(x)$ of weight $d$, 
which correspond to the deformations of the complex structure of $X$. 
 More precisely, $R^Q$ consists of elements of degree $0,d,2d$ and $3d$, and the  dimensions
 of the corresponding subspaces are $1,h^{\mathrm{poly}}_{21},h^{\mathrm{poly}}_{21}$
 and $1$. 
 This degree grading defines a Hodge structure on $R^Q$. As mentioned above 
$R^Q$ is isomorphic to a subgroup of $H^3(X)$. This isomorphism  sends the degree filtration to the Hodge filtration on $H^3(X)$ \cite{Candelas}.
Fermat polynomials have a nice property that there is a bigger symmetry group
  $\prod_i\ZZ_{d/k_i}$ that diagonally acts 
   on $\CC^5$: $\alpha \cdot (x_1, \cdots, x_5) = (\alpha_1^{k_1} x_1, \cdots, \alpha_5^{k_5} x_5), \; \alpha_i^{d/k_i}=1$.
 This action preserves $W_0 = \sum_{i=1}^{5}x_i^{ \frac{d}{k_i}}$. \\
\indent  In particular, the quantum symmetry $Q=\ZZ_d $ is the diagonal   subgroup of  $\prod_i\ZZ_{d/k_i}$.
 The monomial basis $\{e_{\mu}(x) = e_{(\mu_1, \cdots, \mu_5)}(x) = \prod_i x^{\mu_i} \}$ 
of $R^Q$ is an eigenbasis of the phase symmetry  $\ZZ^5_d$, and
each $e_{\mu}(x)$ has a unique weight.
 We can extend the phase symmetry action to the 
parameter space $\{\phi_s\}_{s=1}^{h^{\mathrm{poly}}_{2,1}}$ such that
 $W(x, \phi)$ is invariant 
under this action. That is, if $\alpha \cdot e_{s}(x) = \lambda_{\alpha} e_{s}(x)$ 
for some root of unity $\lambda_{\alpha}$,  then
we must define $\alpha \cdot \phi_{s} := \lambda_{\alpha}^{-1} \phi_s$. In 
particular, the  equations $W(x, \phi) = 0$ and $W(x, \alpha \cdot \phi) = 0$ define the same
CY manifold  because the action of $\alpha$ can be undone by a coordinate 
tranformation of the variables $x_i$. This  means that for Fermat polynomials $W_0$, the
point $\phi = 0$ is an \textit{orbifold} point in the CY moduli space. 
Such a symmetry allows  simplifing the computations significantly.

The  phase symmetry group action obviously preserves the Hodge decomposition.
 The complex conjugation  acts 
on $H^3(X)$ such that $\overline{H^{p,q}(X)} = H^{q,p}(X)$, in particular,
 $\overline{H^{2,1}(X)} = H^{1,2}(X)$.
 Through the isomorphism  between
$R^Q$ and $H^3(X)$, the complex conjugation  also acts  on the elements of the
ring $R^Q. $
%%%%%%%%%%%%%%%%%%%%%%%%%%%%%%%%%%%%%%%%%%%%%%%%%%%%%%%%%%%%%%%%%%%%%%%%%%%%%%%%%%%%%%%
%%%%%%%%%%%%%%%%%%%%%%%%%%%%%%%%%%%%%%%%%%%%%%%%%%%%%%%%%%%%%%%%%%%%%%%%%%%%%%%%%%%%%%%
%%%%%%%%%%%%%%%%%%%%%%%%%%%%%%%%%%%%%%%%%%%%%%%%%%%%%%%%%%%%%%%%%%%%%%%%%%%%%%%%%%%%%%%
 \subsection{Oscillatory representation and computing the periods.}
We introduce the special  bases $\Gamma^{\pm}_{\mu}$ in the homology groups  
$\mathscr{H}_5^{\pm,inv}$ by requiring  their duality to the bases 
in  $H^5_{D_{\pm}}(\CC^5)_{inv}$:
\be
\int_{\Gamma^{\pm}_{\mu}}
 e_{\nu}(x) e^{\mp W_0(x)} \dd^5 x = \delta_{\mu\nu}
\ee
with the corresponding periods
\ba 
&\sigma_{\alpha\mu}^{\pm}(\phi) := \int_{\Gamma^{\pm}_{\mu}}
 e_{\alpha}(x) e^{\mp W(x, \phi)} \dd^5 x, \\
 &\sigma_{\mu}^{\pm}(\phi) := \sigma_{0\mu}^{\pm}(\phi).
\ea
These periods are eigenfunctions of the phase symmetry group action.\\
%%%%%%%%%%%%%%%%%%%%%%%%%%%%%%%%%%%%%%%%%%%%%%%%%%%%%%%%%%%%%%%%%%%%%%%%%%%%%%%%%%%%%%%
 \indent To explicitly compute $\sigma^{\pm}_{\mu}(\phi)$, following \cite{BB, AKBA} we first  expand the exponent in the
 integral regarding the terms in  $W(x,\phi) = W_0(x) + \sum_s \phi_s e_s(x)$,  which are proportional 
$ \phi$, as a perturbation. We then obtain
\be
\sigma^{\pm}_{\mu}(\phi) = \sum_m  \int_{\Gamma^{\pm}_{\mu}} 
 \prod_r e_{r}(x)^{m_r} \, e^{\mp W_0(x)} \, \dd^5 x \; \left(\prod_s\frac{(\pm\phi_s)^{m_s}}{ m_s!}\right),
\ee
where $m := \{m_s \}_{s}, \; m_s \ge 0$, denotes a multi-index of powers of $\phi_s$ in the above expansion.
Because  $\sigma^-_{\mu}(\phi) = (-1)^{|\mu|}\sigma^+_{\mu}(\phi),$
 we focus on $\sigma_{\mu}(\phi) := \sigma^{+}_{\mu}(\phi).$\\
\indent For each of the summands,  the form  $\prod_s e_{s}(x)^{m_s} \, \dd^5 x$
 belongs to $H^5_{D_{\pm}}(\CC^5)_{inv},$ because it is killed by $D_+$ and is $Q$-invariant.
The oscilatory integrals of $D_+$-exact terms are zero,  and therefore 
\be
\int_{\Gamma^{+}_{\mu}} e^{- W_0(x)}
 P(x) \, \dd^5 x   = \int_{\Gamma^{+}_{\mu}} 
 e^{- W_0(x)} (P(x) \,  \dd^5 x + D_+ U ) 
\ee
for any polynomial $P(x)$ and  any polynomial $4-$form $U$.

%%%%%%%%%%%%%%%%%%%%%%%%%%%%%%%%%%%%%%%%%%%%%%%%%%%%%%%%%%%%%%%%%%%%%%%%%%%%%%%%%%%%%%%
%%%%%%%%%%%%%%%%%%%%%%%%%%%%%%%%%%%%%%%%%%%%%%%%%%%%%%%%%%%%%%%%%%%%%%%%%%%%%%%%%%%%%%%

We set  $m_s s_i =  \nu_i + n_i \frac{d}{k_i}  , \; \nu_i < \frac{d}{k_i}$, for later convenience.\\  
To compute 
\be
\int_{\Gamma^{+}_{\mu}} e^{- W_0(x)} \prod_i x_i^{\nu_i + n_i \frac{d}{k_i} } \, \dd^5 x,
\ee
we use the  above trick with
%\be
%\ee
\ba
&\prod_i x_i^{\nu_i + n_i \frac{d}{k_i} } \, \dd^5 x =
&=(-1)\left(n_1-1+\frac{k_1(\nu_1+1)}{d}\right) 
x^{\nu_1 + (n_1 - 1) \frac{d}{k_1} } \prod_{i>1} x_i^{\nu_i + n_i \frac{d}{k_i}} \, \dd^5 x+ D_{+} U .
\ea

where
\be
U=\frac{k_1}{d}x_1^{\nu_1 +1 + (n_1 - 1) \frac{d}{k_1}} \,\prod_{i>1} x_i^{\nu_i + n_i \frac{d}{k_i}} 
\, \dd x_2 \wedge \cdots \wedge \dd x_5.
\ee
%%%%%%%%%%%%%%%%%%%%%%%%%%%%%%%%%%%%%%%%%%%%%%%%%%%%%%%%%%%%%%%%%%%%%%%%%%%%%%%%%%%%%%%
%%%%%%%%%%%%%%%%%%%%%%%%%%%%%%%%%%%%%%%%%%%%%%%%%%%%%%%%%%%%%%%%%%%%%%%%%%%%%%%%%%%%%%%
We  continue this procedure by induction with respect to all $n_i$.\\
We can write the final result  compactly  using Pochhammer's symbols:
\be
\prod_i x_i^{\nu_i + n_i \frac{d}{k_i}} \, \dd^5 x = 
(-1)^{\sum_i n_i}\prod_i \left(\frac{k_i(\nu_i+1)}{d}\right)_{n_i}
\prod_i x_i^{\nu_i} \, \dd^5 x, \; \nu_i < \frac{d}{k_i}.
\ee
where $(a)_{n}=\Gamma(a+n)/\Gamma(a)$.

If any $\nu_i = d/k_i-1$, then the differential form is exact, and
 the integral is zero.\\
 Otherwise, the RHS of the equation is proportional to $e_{\nu}(x)$, and
 we can  use the definition of $\Gamma^+_{\mu}$:
\be
\int_{\Gamma^+_{\mu}} e_{\nu}(x) \, e^{-W_0(x)} \, \dd^5 x = \delta_{\mu\nu}.
\ee 
Doing in this way  and integrating over $\Gamma^+_{\mu}$ we obtain the explicit 
expression for the periods
\be 
\sigma_{\mu}(\phi)=
 \sum_{n_i\ge 0} \prod_i \frac{\Gamma\left(n_i+\frac{k_i(\mu_i+1)}{d}\right)}{\Gamma\left(\frac{k_i(\mu_i+1)}{d}\right)} 
\sum_{m \in \Sigma_n } \prod_s\frac{\phi_s^{m_s}}{ m_s!},
\ee
where 
\be
\Sigma_n = \{m_s\;|\;\sum_s m_s s_i = \mu_i+\frac{d}{k_i}n_i\}.
\ee

%%%%%%%%%%%%%%%%%%%%%%%%%%%%%%%%%%%%%%%%%%%%%%%%%%%%%%%%%%%%%%%%%%%%%%%%%%%%%%%%%%%%%%%
%%%%%%%%%%%%%%%%%%%%%%%%%%%%%%%%%%%%%%%%%%%%%%%%%%%%%%%%%%%%%%%%%%%%%%%%%%%%%%%%%%%%%%
%%%%%%%%%%%%%%%%%%%%%%%%%%%%%%%%%%%%%%%%%%%%%%%%%%%%%%%%%%%%%%%%%%%%%%%%%%%%%%%%%%%%%%%
%%%%%%%%%%%%%%%%%%%%%%%%%%%%%%%%%%%%%%%%%%%%%%%%%%%%%%%%%%%%%%%%%%%%%%%%%%%%%%%%%%%%%%

\subsection{ Computing the antiholomorphic involution $\bf{M_{\mu \nu}}$.}

 We now want to  compute the antiholomorphic involution $M_{\mu \nu}$. For this, we use its connection   with the transition matrix $T$. We must first choose an  real basis of cycles. We  choose Lefschetz thimbles  $L^{\pm}_{\mu}$  as the basis of such cycles.
 The  matrix   $T$ can then  be found ast the transition  matrix that connects  the cycles  $\Gamma^{\pm}_{\mu}$  and Lefschetz thimbles $L^{\pm}_{\mu}$:
$$ \Gamma^{\pm}_{\mu}=(T^{-1})_{\mu \nu}\, L^{\pm}_{\nu}.$$
It follows that   the  transition  matrix $T_{\mu \nu}$ is just given by the integral
$$
T_{\mu \nu} =\int_{L^{\pm}_{\mu}} e_{\nu}(x) \, e^{\mp W_0(x)} \dd^5 x.
$$
After computing this integral,  we obtain the matrix $M$ from the formula  $$M = T^{-1} \bar{T}.$$
%%%%%%%%%%%%%%%%%%%%%%%%%%%%%%%%%%%%%%%%%%%%%%%%%%%%%%%%%%%%%%%%%%%%%%%%%%%%%%%%%%%%%%%
%%%%%%%%%%%%%%%%%%%%%%%%%%%%%%%%%%%%%%%%%%%%%%%%%%%%%%%%%%%%%%%%%%%%%%%%%%%%%%%%%%%%%%%
The Lefschetz thimbles  $L^{\pm}_{\mu}$  are products of one-dimensional cycles $C_{\mu_i}$,
$$
L^{+}_{\mu}   =\prod^5_{i=1}C_{\mu_i},
$$
and  $C_{\mu_i} =  \hat\rho_i^{\mu_i} \cdot C_i$ with $\rho_i = 
e^{\frac {2\pi i k_i}{d}}$.
This definition of the one-dimensional  cycle   $C_{\alpha_i}$ means that this cycle 
is the path in the $x_i$ plane obtained  by the operation $\hat\rho^{\mu_i} $ of  rotating counterclockwise through an  angle   $\frac {2\pi k_i \mu_i}{d}$
from the basic path  $C_i$  depicted on the figure

\begin{figure}[h!]
\centering
\def\svgwidth{4cm}
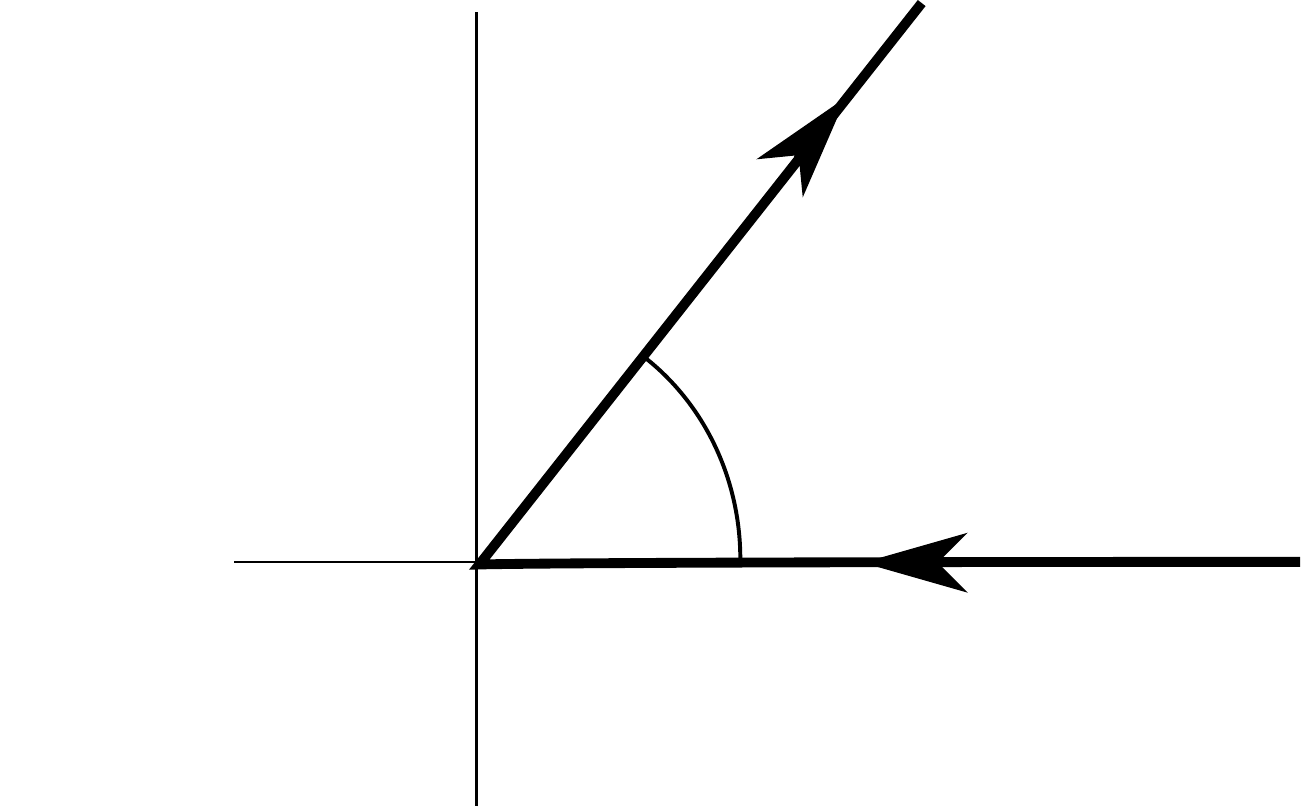
%\includegraphics[width=5cm]{hexagons.png}
%\caption{Basic one-dimensional contour}
\label{fig:c0}
\end{figure}
 By construction, $L^{\pm}_{\mu} $ are the steepest descent/acsent cycles for $\mathrm{Re}W_0$. We now compute  $T_{\alpha\mu}$ explicitly 
\be
T_{\alpha\mu}= \int_{L^+_{\alpha}} e_{\mu} \, e^{-W_0} \, \dd^5 x = 
\rho^{(\bar{\alpha},\bar{\mu})} A(\mu),
\ee
where $A_{\mu}$ is a product of five gamma integrals,
\be
A_{\mu} =\prod_i \left( \frac{k_i}{d}\right) \Gamma\left(\frac{k_i(\mu_i+1)}{d} \right).
\ee
Then
\be
T^{-1}_{\bar{\mu}\bar{\alpha}} = B(\mu)[\bar{\rho}^{(\bar{\mu}+1, \bar{\alpha})}-1],
\ee
\be
B(\mu) = \prod_i \frac{1}{\Gamma\left(\frac{k_i(\mu_i+1)}{d}\right)},
\ee

 $$ M_{\mu\nu} =  (T^{-1} \bar{T})_{\mu\nu} = 
\prod_i \gamma\left(\frac{k_i(\mu_i+1)}{d} \right) \delta_{\mu, \rho-\nu},$$

$$ \gamma(x) = \frac{\Gamma(x)}{\Gamma(1-x)}.$$
%%%%%%%%%%%%%%%%%%%%%%%%%%%%%%%%%%%%%%%%%%%%%%%%%%%%
%%%%%%%%%%%%%%%%%%%%%%%%%%%%%%%%%%%%%%%%%%%%%%%%%%%%%
\subsection {K\"ahler  potential  for the  moduli space of Fermat threefolds.}
\label{sec:res}

Subtitung the explicit expressions for the periods $\sigma_\mu$, the pairing $\eta_{\mu \nu}$, and the anti-involution $M$ in the above expression for  the   K\"ahler potential  on the   moduli space, we obtain
  
\be \label{FermatKah}
e^{-K(\phi)} = \sum_{\mu}^{}(-1)^{\deg (\mu)/d} \prod_i{\gamma\left(\frac{k_i(\mu_i+1)}{d}\right)}
 |\sigma_{\mu}(\phi)|^2, 
\ee
where index $\mu$ in the summation runs through the whole invariant Milnor
ring basis
$0 \leq \mu_i \leq \frac{d}{k_i}-2$, \quad $\sum_{i=1}^{5}\mu_i=0,d,2d,3d $,

\be \label{FermatPer}
\sigma_{\bf{\mu}}(\phi)=
 \sum_{n_1,...,n_5\ge 0} \prod_{i=1}^{5}
\frac{\Gamma(\frac{k_i(\mu_i+1)}{d}+n_i)}{\Gamma(\frac{k_i(\mu_i+1)}{d})}
\sum_{m \in \Sigma_n } \prod_s \frac{\phi_{s}^{m_{\bf{s}}}} { m_s!} ,
\ee
and
\be
  \gamma(x) = \frac{\Gamma(x)}{\Gamma(1-x)}, \qquad
\Sigma_n = \{m_{\bf{s}} \;|\;\sum_s m_{\bf{s}} s_i = \mu_i + \frac{d}{k_i}n_i\}.
\ee

\section{Conclusion.}

We have applied our method  to the CY
hypersurfaces given as zero sections of Fermat polynomials in weighted projective
spaces. We use a better way to compute the real (and even integral structure) for
the periods compared with our previous work.

While preparing  this paper we learned that the periods and their
integral structure were basically computed in a different language
 and different setting in the mathematical literature~\cite{IMR, Iritani}.

The  possible application of the method, used in this paper, which should be done,  
is the computation  for the invertible  singularities of the Berglund--Hubsch type 
\cite{BerHub1, BerHub2}. 

It would   be also interesting to know  connections with the other points of moduli spaces, that is our method gives Special Geometry metric as a power series around orbifold
points of the moduli spaces, which correspond to nonsingular CY manifolds. However, 
there are many interestiong points in the moduli space of CY varieties, many of which
are singular such as maximal unipotent monodromy points (mirror to large volume points describing Gromov-Witten theory of the mirror manifold)
or conifold points which are the simplest degenerations of the CY manifold.

We are grateful to V. Batyrev, B. Everett, D. Gepner, H. Iritani, S. Katz and V. Vasiliev for the  useful discussions.
K.A. is grateful to  the Mathematical Sciences Research Institute in Berkeley for hospitality.

The work  is supported by the Russian Science Foundation under grant 18-12-00439 and performed in Landau Institute for Theoretical Physics.

%\appendix

%%%%%%%%%%%%%%%%%%%%%%%%%%%%%%%%%%%%%%%%%%%%%%%%%%%%%%
%%%%%%%%%%%%%%%%%%%%%%%%%%%%%%%%%%%%%%%%%%%%%%%%%%%%%%%%%%%%%%%%%%%%%%%%%%%%%%%%%%%%%%%
 
%%%%%%%%%%%%%%%%%%%%%%%%%%%%%%%%%%%%%%%%%%%%%%%%%%%%%%%%%%%%%%%%%%%%%%%%%%%%%%%%%%%%%%%
%%%%%%%%%%%%%%%%%%%%%%%%%%%%%%%%%%%%%%%%%%%%%%%%%%%%%%%%%%%%%%%%%%%%%%%%%%%%%%%%%%%%%%%

%%%%%%%%%%%%%%%%%%%%%%%%%%%%%%%%%%%%%%%%%%%%%%%%%%%%
%\bibliographystyle{JHEP}
%\bibliography{cymoduli}
\printbibliography

\end{document}